\documentclass[preprint,showpacs,preprintnumbers,amsmath,amssymb]{revtex4}

\usepackage{dcolumn}
\usepackage{bm}
\input epsf
\usepackage{color}

\newcommand{\be}{\begin{equation}}
\newcommand{\en}{\end{equation}}

\begin{document}


\title{Constraints on Holographic cosmologies from strong lensing systems}

\author{V\'ictor H. C\'ardenas}
\email{victor.cardenas@uv.cl}
\author{Alexander Bonilla}
\email{alex.bonilla@uv.cl}
\author{Ver\'onica Motta}
 \email{veronica.motta@uv.cl}
\affiliation{ Departamento de F\'isica y Astronom\'ia, Facultad de
Ciencias, Universidad de Valpara\'iso, Gran Breta\~na 1111,
Valpara\'iso, Chile }

\author{Sergio del Campo}
\email{sdelcamp@ucv.cl} \affiliation{Instituto de F\'isica,
Pontificia Universidad Cat\'olica de Valpara\'iso, Av. Brasil 2950,
Valpara\'iso, Chile.}

\date{\today}

\begin{abstract}
We use strongly gravitationally lensed (SGL) systems to put
additional constraints on a set of holographic dark energy models.
Data available in the literature (redshift and velocity dispersion)
is used to obtain the Einstein radius and compare it with model
predictions. We found that the $\Lambda$CDM is the best fit to the
data. Although a preliminary statistical analysis seems to indicate
that two of the holographic models studied show interesting
agreement with observations, a stringent test lead us to the result
that neither of the holographic models are competitive with the
$\Lambda$CDM. These results highlight the importance of Strong
Lensing measurements to provide additional observational constraints
to alternative cosmological models, which are necessary to shed some
light into the dark universe.
\end{abstract}

\maketitle

\section{Introduction}

The accelerating expansion of the Universe is a fundamental
challenge to both particle physics and cosmology. Although initially
the evidence emerge from studies of Type Ia supernova (SNIa)
\cite{01}, now we have strong indications from probes like the large
scale structure (LSS, \cite{02}), cosmic microwave background (CMB,
\cite{03}), the integrated Sachs--Wolfe effect (ISW, \cite{isw}),
baryonic acoustic oscillations (BAO, \cite{Eisenstein:2005su}) and
gravitational lensing \cite{weakl}. The source of this mysterious
cosmic acceleration is dubbed dark energy (DE). The simplest
candidate is a cosmological constant $\Lambda$, which leads to the
successful $\Lambda$-cold-dark-matter ($\Lambda$CDM) model. Although
it fits most of the observational data rather well, it suffers from
two main problems, namely: the low value of the vacuum energy and
the coincidence problem \cite{1}. To address these two problems, the
cosmological constant is replaced by a time varying quantity,
leading to the dynamical DE models. The most studied models are
scalar field ones which comprehend, e.g., quintessence \cite{2},
K-essence \cite{3} and tachyon fields \cite{4}. Usually, dark matter
(DM) and DE are assumed to evolve independently, however, there is
no reason to neglect interactions in the dark sector
\cite{interaction}. Because both dark components are characterized
through their gravitational effects, it is natural to consider
unified models of the cosmological substratum in which one single
component plays the role of DM and DE simultaneously. Examples of
this type of models are the Chaplygin gas \cite{5}, and bulk-viscous
models \cite{bv}.

Among the many approaches to describe the dark components, the
holographic DE models have received considerable attention
\cite{6,cohen}. The holographic dark energy is one of the emergent
dynamical DE model proposed in the context of fundamental principle
of quantum gravity, so called holographic principle. This principle
arose from black hole and string theories. The holographic principle
states the number of degrees of freedom of a physical system, apart
from being constrained by an infrared cutoff, it should be finite
and it should scale with its bounding area rather than with its
volume\cite{7}. Specifically, it is derived with the help of
entropy-area relation of thermodynamics of black hole horizons in
general relativity which is also known as the Bekenstein-Hawking
entropy bound, i.e., $S \simeq M_p^2 L^2$, where $S$ is the maximum
entropy of the system of length $L$ and $M_p=1/\sqrt{8\pi\,G}$ is
the reduced Planck mass. In general terms, the inclusion of the
holographic principle into cosmology, makes possible to find the
upper bound of the entropy contained in the universe\cite{FS98}.

Using this idea it was possible to obtain a theoretical relation
between a short distance (ultraviolet) cutoff and a long distance
(infrared, IR) cutoff \cite{cohen}. Considering $L$ as a cosmological
length scale, different choices of this cutoff scale results in
different DE models. For example, when identifying L with the Hubble
radius $H^{-1}$, the resulting DE density turns out to be very
close to the observed critical energy density \cite{cohen}.
 Li \cite{li} studied the use of both the particle and event
horizons as the IR cutoff length. He found that apparently only a
future event horizon cutoff can give a viable DE model. 
More recently, it was proposed a new cutoff scale, given by the
Ricci scalar curvature \cite{021,022}, resulting in the so-called
holographic Ricci DE models. Thus, in general terms, the holographic
DE model suffers the IR-cutoff choice problem. On the other hand,
holographic DE model have been tested and constrained by various
astronomical observations \cite{obs}, in some cases also including
spatial curvature \cite{cont}. A special class are those models in
which the holographic DE is allowed to interact with
DM\cite{8,9,10,HDE}.

In this article we use strongly gravitationally lensed (SGL)
systems, to provide additional constraints on these holographic DE
models. The idea of using such systems was discussed first in
\cite{Futamase:2001a} and also in \cite{biesiada2006}. We use the
data set first used in \cite{Cao:2011bg} (see also
\cite{Liao:2012ws}) consisting in 70 data points from Sloan Lens ACS
(SLACS), and galaxy clusters from optical and X-ray surveys
\cite{ota2004,yu2011}.

Some of the holographic DE models we chose corresponds to those
first discussed in \cite{delCampo:2011jp} under the constraint of
various astronomical observations, such that SNIa and from the
history of the Hubble parameter. Using SGL features we compared
three cases of interacting DE models, and studied the relation
between the energy density ratio of DM and DE and the
equation-of-state (EoS) parameter in these cases. An interesting
result of this study is that the role of potential interactions in
the dark sector could be clarified. It is noteworthy that any
interaction model introduces relations between the matter content
and the EoS.


This paper is organized as follows: In Section II we describe the
data considered in this work to put in tension our theoretical
models. The latter are described in section III, the results are
displayed in section IV, and the conclusions in section V.

\section{The Sample}

The discovery of the first lensed quasar Q0957+561\cite{Wetal79}
brings an interesting possibility to use SGL systems as cosmological
tools \cite{refsdal1966,blandford1992}. Some of the statistical
methods which use strong lensing to constraint cosmological models
are for example: (i) The probability of strong lensing event is
known to be sensitive to dark energy
\cite{turner1990},\cite{fukujita1990}. Recent results are in
agreement with LCDM cosmological model e.g., \cite{Chae:2002uf},
\cite{Mitchell:2004gw}, \cite{Oguri:2008ci}. (ii) The differential
probability distribution of lens redshifts is fairly insensitive to
both the source quasar population and magnification bias
\cite{Kochanek1992}, \cite{Ofek:2003sp}, \cite{Chae:2003zv},
\cite{Matsumoto:2007vc}, \cite{Chae:2008ni}, \cite{Cao:2011aq}.
Unfortunately, small number statistics remain a significant
limitation for the cosmological results.

Lensing phenomena are sensitive to the geometry of the cosmological
background since the appearance of an image depends on the distances
between source, lens and observer \cite{B10}. By having information
about these distances from observations (using redshift
measurements)  we will be able to constrain cosmological models.

The advantage of this method is that it is independent of the Hubble
constant value and is not affected by dust absorption or source
evolutions (e.g., as SNIa \cite{riess1996}). However, it depends on
the measurements of $\sigma$ and lens modeling (e.g. singular
isothermal sphere (SIS) or singular isothermal ellipsoid (SIE)
assumption).

Hundreds of lens systems are being discovered in ongoing surveys
(Herschel Lensing Survey, \cite{HLS}, South pole telescope,
\cite{Hezaveh:2013jwa}) and in the next decade 10000 systems are
expected to be detected with the Large Synoptic Survey Telescope and
Euclid \cite{Treu:2013rpx}. With such a huge number of data, and
larger redshift coverage, SGL will provide a level of precision in
cosmology higher than other technics.

In what concern to data manipulation, and following
\cite{Cao:2011bg}, we selected 59 strong lens systems from the Sloan
Lens ACS Survey (SLACS, \cite{bolton2008}) and the Lens Structure
and Dynamic survey (LSD, \cite{koopmans2002,koopmans2003}), and 11
from CfA-Arizona Space Telescope LEns Survey \cite{castles}. SLACS
systems where selected from the Sloan Digital Sky Survey (SDSS)
based on the presence of absorption-dominated galaxy continuum at
one redshift and emission lines at another higher redshift
(\cite{bolton2004}). CASTLES obtained HST images for known
galaxy-mass gravitational lens systems.

It has been shown the singular isothermal sphere (SIS) is an
accurate first-order approximation for an elliptical galaxy acting
as lens (
\cite{kochanek1996,king1997,fassnacht1998,rusin2005,koopmans2006,koopmans2009,treu2006a,treu2006b}).
In these cases, the modeled SIS velocity dispersion ($\sigma_{mod}$)
is in good agreement with the central velocity dispersion measured
spectroscopically
($\sigma_{obs}$)(\cite{treu2006a,treu2006b,grillo2008,koopmans2009}).
The Einstein ring radius is defined as \be \theta_E=4\pi
\frac{D_A(z_L,z_S)}{D_A(0,z_S)}
\frac{\sigma^2}{\overline{c}^2}\label{Anillo} \en where
$\overline{c}$ represents the speed of light and $\sigma$ refers to
the central velocity dispersion  observed ($\sigma_{obs}$) or
modeled ($\sigma_{mod}$).

Based on observations of  X-ray, there is a strong indication that
DM halos are dynamically hotter than the luminous stars, then the
velocity dispersion $\sigma_{SIS}$ of the SIS model is different
from the observed stellar velocity dispersion $\sigma_{obs}$
\cite{white1998}. The authors of \cite{Cao:2011bg} and
\cite{wang2012} have used a SIS model and an extra factor $f_{E}$ to
account for: (i) systematical errors in the observed velocity
dispersion, (ii) differences between $\theta_{E}$ obtained from SIS
and SIE and image separation.

However, \cite{treu2006a} has used the SLACS lenses to compare the
central velocity dispersions with the best SIE lensing model. They
found a factor $f=\sigma_{0}/\sigma_{SIE}=1.01 \pm0.02$, with
$0.065$ rms scatter. As the rms error expected observationally is
less than the $7 \%$, which is by far less than the error in other
parameters (which are around $20 \%$), we prefer not to add a new
parameter in our analysis.

Here, $D_A(z_L,z_S)$ and $D_A(0,z_S)$ represent the angular diameter
distance between lens and source and between observer and lens,
respectively and $z_L$ and $z_S$ corresponds to the lens and source
redshifts respectively. The ratio between these two angular diameter
distances constraint cosmological models, since this distance in a
flat FRW metric corresponds to
\be D_A(z,{\bf p}) = \frac{\overline{c}}{H_0(1+z)}\int_0^z
\frac{dz'}{E(z';{\bf p})},\label{Da} \en
for a given $z$. The parameter ${\bf p}$ specifies the set of
cosmological parameters that enter into the model, $H_0$ is the
current value for the Hubble parameter. The function $E(z,{\bf p})$
represents the dimensionless expansion rate and it is obtained from
the Friedmann Equation $H^2 \sim \rho$, where $\rho$ represents the
total energy density, via the ratio \be E(z,{\bf p}) \equiv
\frac{H(z,{\bf p})}{H_0}. \label{E}\en

\section{The models}

In a flat FRW metric we consider the universe to be composed by
pressureless matter, $\rho_{m}$, and an holographic dark energy
component $\rho_{H}$. The Friedmann equation in this case becomes
\begin{equation}
3 H^2 = 8\pi\, G (\rho_{m} + \rho_{H})  \  .\label{Fried1}
\end{equation}
Allowing the components to interact, we can write
\begin{equation}
\dot{\rho}_{m} + 3 H \rho_{m}  = Q  = -\dot{\rho}_{H} - 3 H
(\rho_{H} + p_H).\label{cons1}
\end{equation}
where $Q$ is an interaction term which can be an arbitrary function
of the Hubble parameter $H$ and the energy densities $\rho_m$ and
$\rho_H$, and $p_H$ represents the pressure related to the
holographic part. Here, the EoS parameter $w$ is defined as the
ratio $p_H/\rho_H$.

By introducing the ratio $r$ between $\rho_{m}$ and $\rho_{H}$ so
that $r= \rho_{m}/\rho_{H}$, we obtain the Hubble rate changes as

\begin{equation}
\frac{d\ln H}{d\ln a}  = - \frac{3}{2}\left(1 + \frac{w(a)}{1 +
r(a)}\right) \ . \label{dH}
\end{equation}
The parameter $r$ changes as 
\begin{equation}
\dot{r} = 3Hr\,\left(1 + r\right)\,\left[\frac{w}{1+r} +
\frac{Q}{3H\rho_{m}}\right]\ , \label{dr2}
\end{equation}
where equation (\ref{cons1}) was used.

We may write the holographic energy density as \cite{cohen,li}
\begin{equation}
\rho_H=\frac{3\,c^2\,M_p^2}{L^{2}} \ ,\label{ans}
\end{equation}
where $L$ represents the IR cutoff scale and $M_p$ is the reduced
Planck mass introduced previously (an arbitrary constant). In the
holographic DE model it is assumed that the energy in a given box
should not exceed the energy of a black hole of the same size. This
means that \be L^3\rho_H \leq M_p^2 L. \en In this context the
numerical constant $c^{2}$ is related with the degree of saturation
of the previous expression.


Next we need to identify $L$ with a cosmological length scale. In
the literature, three choices of $L$ have been considered: the
Hubble scale, the future event horizon, and a scale proportional to
the inverse square root of the Ricci scalar. Each of these cases
will be analyzed in the subsequent sections.

Using (\ref{ans}) and (\ref{cons1}) it is easy to see that
\begin{equation}
\Gamma \equiv \frac{Q}{\rho_{H}} = 2\frac{\dot{L}}{L} -
3H\left(1+w\right) \ , \label{QL}
\end{equation}
where $\Gamma$ corresponds to the rate by which $\rho_{H}$ changes
as a result of the interaction $Q$. In this expression, for $Q=0$
there is a specific relationship between $w$ and the ratio of the
rates $\dot{L}/{L}$ and $H$. Of course, any non-vanishing $Q$ will
modify this relationship.

From expressions (\ref{QL}) and (\ref{dr2}) it is found that the
energy density ratio $r$ evolves as
\begin{equation}
\dot{r} = - 3H\,\left(1 + r\right)\,\left[1 + \frac{w}{1+r} -
\frac{2}{3} \frac{\dot{L}}{H L}\right]\ . \label{drL}
\end{equation}
Note that, in general, different choices of the cutoff scale $L$
result in different relations between $w$ and $r$. It will turns out
that for a Hubble-scale cutoff the ratio $r$ is necessarily
constant. For the other two choices, the future event horizon and
the Ricci-scale cutoffs, the relationships between $w$ and $r$ vary
with time. In particular, in both these cases a constant ratio $r$
requires a constant EoS parameter $w$. In the following sections we
study the three choices for $L$.


\section{Results}

In this section we perform the statistical analysis of each of the
holographic models discussed in the previous section. The analytical
derivation of each model was presented in \cite{delCampo:2011jp} and
here we only display a summary of the results.

In all the plots below, we explicitly show the 1$\sigma$ and
2$\sigma$ confidence contours, with (continues line) and without
(dashed lines) the strong lensing data.

In order to probe the above models against observations, we consider
four background tests which are directly related to the behavior of
the function $H(z)$, i.e. the Hubble parameter as a function of the
redshift $z$: SGL systems
\cite{Futamase:2001a},\cite{biesiada2006},\cite{Cao:2011bg},
supernova type Ia \cite{Union2}, massive and passively evolving
early-type galaxies as ``cosmic chronometers" \cite{Jimenez:2001gg},
and other technics, which gives a direct measure of the $H(z)$
function \cite{Farooq:2013hq}, and the baryonic acoustic
oscillations \cite{Eisenstein:2005su}. We shall present the results
for a combined analysis of these four tests. Because all the models
we consider, describe the universe evolution from the matter
domination epoch to the onset of cosmic acceleration (where
radiation is negligible) we only use the BAO data points from the
WiggleZ experiment \cite{Blake:2011en}.

The supernova type Ia (SNIa) test is based on the luminosity
distance function
\begin{equation}
D_L = (1 +
z)\frac{\overline{c}}{H_0}\int_0^z\frac{dz}{\sqrt{H(z)}}.
\end{equation}
The observational relevant quantity is the moduli distance, given
by
\begin{equation}
\mu = m - M = 5\ln\biggr(\frac{D_L}{Mpc}\biggl) + 25,
\end{equation}
where $m$ is the apparent magnitude and $M$ is the absolute
magnitude of a given supernova. In what follows we shall use the
data set of the Union2 sample \cite{Union2}.

The BAO measurements considered in our analysis are obtained from
the WiggleZ experiment \cite{Blake:2011en}. The $\chi^2$ for the
WiggleZ BAO data is given by
\begin{equation}
\chi^2_{\scriptscriptstyle WiggleZ} = (\bar{A}_{obs}-\bar{A}_{th})
C_{\scriptscriptstyle WiggleZ}^{-1} (\bar{A}_{obs}-\bar{A}_{th})^T,
\end{equation}
where the data vector is $\bar{A}_{obs} = (0.474,0.442,0.424)$ for
the effective redshift $z=0.44,0.6$ and 0.73. The corresponding
theoretical value $\bar{A}_{th}$ denotes the acoustic parameter
$A(z)$ introduced by \cite{Eisenstein:2005su}:
\begin{equation}
A(z) = \frac{D_V(z)\sqrt{\Omega_{m}H_0^2}}{cz},
\end{equation}
and the distance scale $D_V$ is defined as
\begin{equation}
D_V(z)=\frac{1}{H_0}\left[(1+z)^2D_A(z)^2\frac{cz}{E(z)}\right]^{1/3},
\end{equation}
where $D_A(z)$ is the Hubble-free angular diameter distance which
relates to the Hubble-free luminosity distance through
$D_A(z)=D_L(z)/(1+z)^2$. The inverse covariance
$C_{\scriptscriptstyle WiggleZ}^{-1}$ is given by
\begin{equation}
C_{\scriptscriptstyle WiggleZ}^{-1} = \left(
\begin{array}{ccc}
1040.3 & -807.5 & 336.8\\
-807.5 & 3720.3 & -1551.9\\
336.8 & -1551.9 & 2914.9
\end{array}\right).
\end{equation}

Another test we use is the age of the very old galaxies that have
evolved passively. Our analysis is based on the 28 data points
listed in reference \cite{Farooq:2013hq}. The basic idea under this
approach is based on the measurement of the differential age
evolution as a function of redshift of these galaxies, which
provides a direct estimate of the Hubble parameter $H(z) \simeq
-1/(1 + z)\triangle z/\triangle t$.

The fourth test comes from the formula of the Einstein radius for a
singular isothermal sphere (SIS) model expressed by equation
(\ref{Anillo})
which depends on the ratio of the angular diameter distances
between lens and source and between observer and lens.
In this method, the cosmological information enters into a distance
ratio
\begin{equation}\label{dth}
D^{th}(z_L,z_S)=\frac{\int_0^{z_S}dx/E(x)}{\int_{z_L}^{z_S}dx/E(x)},
\end{equation}
where the function $E$ represents the dimensionless expansion rate
introduced previously in (\ref{E}). The observational data come from
\begin{equation}\label{dobs}
D^{obs}=4\pi \frac{\sigma_{SIS}^2}{\theta_{E}\,\overline{c}^2}.
\end{equation}
For each of these observational tests we evaluate the fitting
function $\chi^2$, given by
\begin{equation}
\chi^2 = \sum_{i=1}^n\frac{(\epsilon^{th}_i -
\epsilon^{ob}_i)^2}{\sigma^2_i},
\end{equation}
where $\epsilon^{th}_i$ stands for a  theoretical estimation of the
$ith$ data of a given quantity (moduli distance $\mu(z)$, parameters
$R$ and $\cal{A}$, $H(z)$), $D^{th}(z_L,z_S)$, and $\epsilon^{ob}_i$
stands for the corresponding observational data, $\sigma_i$ being
the error bar. In the following we will treat the cases separately.

\subsection{Hubble radius}

\begin{figure}[h!]
\centering \leavevmode\epsfysize=8cm \epsfbox{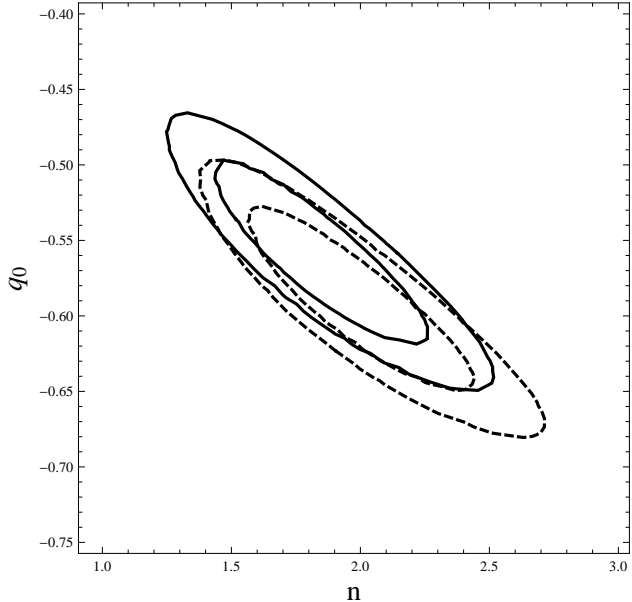}\\
\caption{\label{HR} Here we display the $68.27\% $ and $95.45\%$
confidence regions for the parameters $q_0$ and $n$ of the
Hubble-scale cutoff model of Eq.(\ref{HH}).}
\end{figure}

For $L=H^{-1}$ the holographic DE density is
\begin{equation}
\rho_H= 3\,c^2\,M_p^2 \,H^{2}  \ . \label{rh}
\end{equation}
a power of the Hubble rate, equivalent to
\begin{equation}
\frac{\Gamma}{3 H r} = \mu \left(\frac{H}{H_{0}}\right)^{-n}
\label{ansH}
\end{equation}
The quantity $\mu$ is an interaction constant. Different interaction
rates are characterized by different values of $n$. Considering $n
\neq 0$ we found
\begin{equation}
H(z) = H_0\biggr(\frac{1}{3}\biggl)^{1/n}\biggr[(1 - 2q_0) + 2(1 +
q_0)(1 + z)^\frac{3n}{2}\biggl]^\frac{1}{n}\ . \label{HH}
\end{equation}
The free parameters are $H_{0}$, $q_{0}$ and $n$. In a first step,
the Hubble parameter $H_0$ is determined by minimizing the
three-dimensional $\chi^{2}$ function. The remaining parameters then
are $q_{0}$ and $n$, for which we perform the statistical analysis.
The results are displayed in Figure \ref{HR}
As was mentioned in \cite{delCampo:2011jp} this model for $n=2$ is
similar to the $\Lambda$CDM model.

The best fit value for the parameters are $n=1.82^{+ 0.44}_{-0.39}$
and $q_0=-0.56^{+ 0.06}_{-0.06}$. The dark matter density used was
$\Omega_m=0.25$, and the best fitted value obtained for the Hubble
parameter was $h=0.697$. We use this value for $h$ to obtain the
confidence contour of the parameters displayed in Figure \ref{HR},
showing that at one sigma this model is slightly different from the
$\Lambda$CDM model. However, the $\chi^2_{red}=1.213 $ shows that
this model is among the worst fit in this work although it is a
reasonable fit, it is not as good as the $\Lambda$CDM model
($\chi^2_{red}=1.030$). We also notice the impact of SGL data
points. Considering they are only the $10 \%$ of the whole data set
in this work, the SGL data shift the best fit to a smaller value for
$n$ and towards a slightly higher value for $q_0$.

\subsection{Future event horizon with $\xi = 1$.}

\begin{figure}[h!]
\centering \leavevmode\epsfysize=8cm \epsfbox{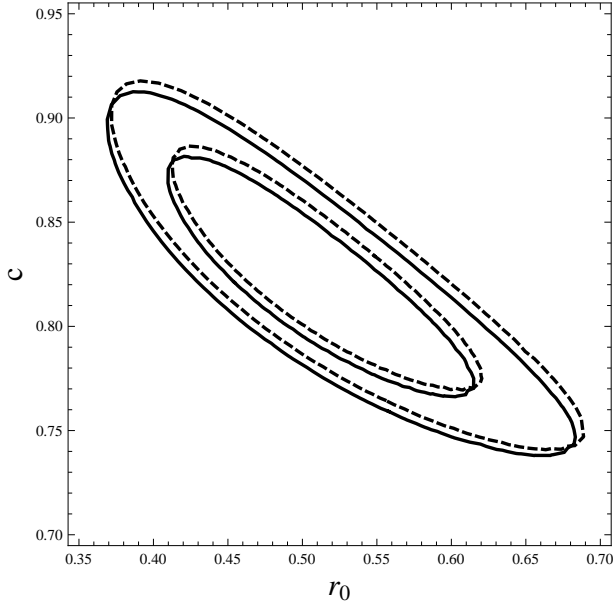}\\
\caption{\label{FEH1} Here we display the $68.27\% $ and $95.45\%$
confidence regions for the parameters $r_0$ and $c$ of the Future
event horizon model  with $\xi=1$ Eq.(\ref{Hev1}).}
\end{figure}

With $L=R_{E}$ the holographic DE density (\ref{ans}) is
\begin{equation}
\rho_H= \frac{3\,c^2\,M_p^2}{R_{E}^{2}}  \  ,\label{rhoE}
\end{equation}
where
\begin{equation}
R_E(t)=a(t)\,\int_t^\infty\,\frac{dt'}{a(t')}=a\,\int_a^\infty\,\frac{da'}{H'\,a'^2}\
\label{re}
\end{equation}
is the future event horizon. The dark-energy balance (\ref{cons1})
can be written as
\begin{equation}
\dot{\rho}_{H} + 3 H (1 + w_{eff})\rho_{H} = 0 \  \label{deeff}
\end{equation}
with an effective EoS
\begin{equation}
w_{eff} = w + \frac{Q}{3H\rho_{H}} = - \frac{1}{3}\left(1 +
\frac{2}{R_{E}H} \right)\ ,
  \label{weff}
\end{equation}
Although this quantity does not directly depend on $w$, the ratio
$r$ that enters $R_{E}H$ is determined by $w$ via
\begin{equation}
\dot{r} = - H\left(1 + r\right) \left[1 + 3\frac{w}{1+r} +
\frac{2}{R_{E}H}\right]\ . \label{drfh}
\end{equation}
Assuming a power-law dependence for the energy-density ratio
\begin{equation}
r = r_{0}a^{-\xi} \ . \label{rxi}
\end{equation}
we can solve the system. As was mentioned in \cite{delCampo:2011jp}
and \cite{dalal}, any value $\xi < 3$ makes the coincidence problem
less severe than in the $\Lambda$CDM model. For this reason we
consider the models with $\xi = 1, 2, 3$ separately. In the first
case, $\xi=1$ we obtain
\begin{equation}
H(z) = H_0(1 + z)^{3/2 - 1/c}\sqrt{\frac{1 + r_0(1 + z)}{r_0 +
1}}\biggr[\frac{\sqrt{r_0(1 + z) + 1} + 1}{\sqrt{r_0 + 1} +
1}\biggl]^{2/c}\ . \label{Hev1}
\end{equation}
Since $\xi$ is fixed, we have $H_{0}$, $r_{0}$ and $c$ as free
parameters. $H_{0}$ is obtained as in the previous case. The
free-parameter space then consists of $r_{0}$ and $c$. The results
are displayed in Figure \ref{FEH1}.

The best fit value we obtained for the Hubble parameter was
$h=0.687$. Using it, the best fit parameters are $r_{0}=0.50
^{+0.11}_{-0.09}$, and $c=0.82^{+0.06}_{-0.06}$. The matter
density is related to $r_0$ through $r_0=\Omega_0/(1-\Omega_0)$
then, our statistical analysis implies
$\Omega_0=0.33^{+0.07}_{-0.06}$. The $\chi^2_{red} = 1.084$
indicates although it is a reasonable fit, being better than the Hubble Radius, it is not as good as the $\Lambda$CDM model ($\chi^2_{red}=1.030$).

In comparison to the results of \cite{delCampo:2011jp}, our best fit
values differs appreciably at one sigma, although at 3$\sigma$
essentially there is no difference.

\subsection{Future event horizon with $\xi = 2$.}

\begin{figure}[h!]
\centering \leavevmode\epsfysize=8cm \epsfbox{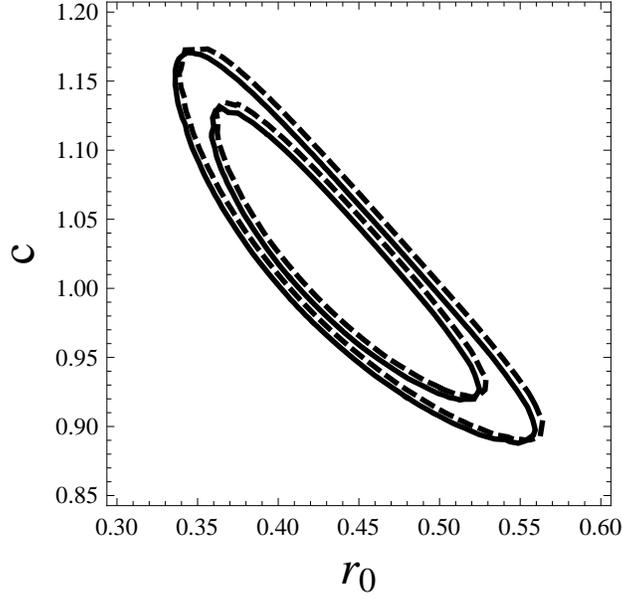}\\
\caption{\label{FEH2} Here we display the $68.27\% $ and $95.45\%$
confidence regions for the parameters $r_0$ and $c$ of the Future
event horizon model  with $\xi=2$ Eq.(\ref{Hev2}).}
\end{figure}

Performing a similar analysis as before, this time with $\xi=2$, we
obtain
\begin{equation}
H(z) = H_0(1 + z)^{1 - 1/c}\sqrt{\frac{1 + r_0(1 + z)^2}{r_0 +
1}}\biggr[\frac{\sqrt{r_0(1 + z)^2 + 1} + 1}{\sqrt{r_0 + 1} +
1}\biggl]^{1/c}\ . \label{Hev2}
\end{equation}
As in the previous case, the free parameters are  $H_{0}$, $r_{0}$
and $c$. The results are displayed in Figure \ref{FEH2}. The best
fit we obtained for the Hubble parameter is $h=0.697$. Using it, the
best fit parameters are $r_{0}=0.43^{+0.09}_{-0.09}$, and
$c=1.02^{+0.11}_{-0.10}$. The matter density implied by our
statistical analysis gives $\Omega_0=0.30^{+0.06}_{-0.05}$. The
$\chi^2_{red} = 1.034$ shows this model is a competitive fit to all
the data compared to the $\Lambda$CDM. In comparison to the results
of \cite{delCampo:2011jp}, our best fit value for $r_0$ is
essentially the same, although our best value for $c$ is slightly
higher at one sigma, at 3$\sigma$ essentially there is no
difference.

\subsection{Future event horizon $\xi = 3$.}

\begin{figure}[h!]
\centering \leavevmode\epsfysize=8cm \epsfbox{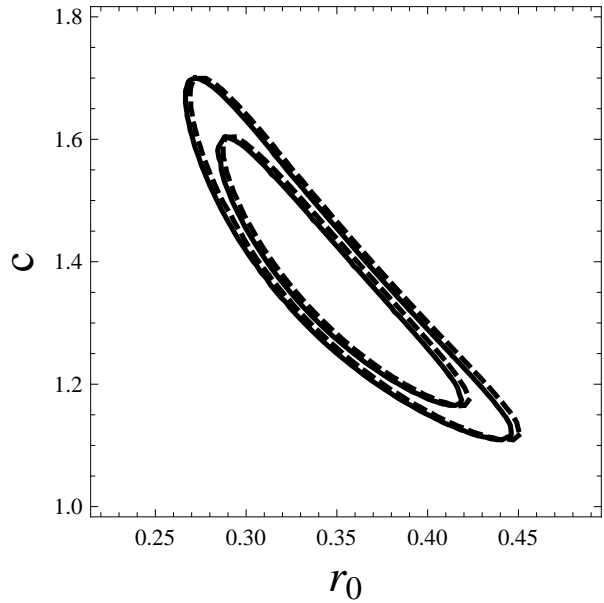}\\
\caption{\label{FEH3} Here we display the $68.27\% $ and $95.45\%$
confidence regions for the parameters $r_0$ and $c$ of the Future
event horizon model  with $\xi=3$ Eq.(\ref{Hev3}).}
\end{figure}

Using this time $\xi=3$, we obtain:
\begin{equation}
H(z) = H_0(1 + z)^{1/2 - 1/c}\sqrt{\frac{1 + r_0(1 + z)^3}{r_0 +
1}}\biggr[\frac{\sqrt{r_0(1 + z)^3 + 1} + 1}{\sqrt{r_0 + 1} +
1}\biggl]^{2/(3c)}\ . \label{Hev3}
\end{equation}
with confidence regions displayed in Fig.(\ref{FEH3}). The best fit
value we obtained for the Hubble parameter is $h=0.71$. The best fit
parameters are $r_{0}=0.35^{+0.07}_{-0.06}$, and
$c=1.36^{+0.24}_{-0.20}$. The matter density implied by our
statistical analysis gives $\Omega_0=0.26^{+0.05}_{-0.05}$. The
$\chi^2_{red} = 1.052$ shows this among the best fit in this work.
In comparison to the results of \cite{delCampo:2011jp}, our best fit
values differs completely even at 3$\sigma$.

\subsection{Ricci scale with CPL parametrization.}

\begin{figure}[h!]
\centering \leavevmode\epsfysize=8cm \epsfbox{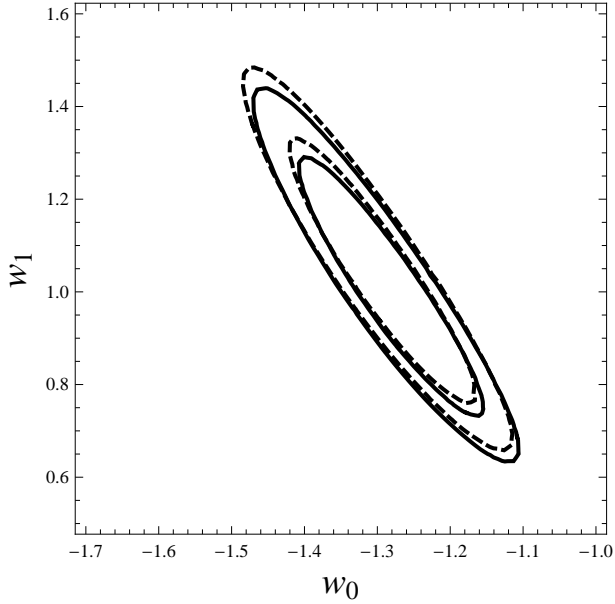}\\
\caption{\label{RS1} Here we display the $68.27\% $ and $95.45\%$
confidence regions for the parameters $w_0$ and $w_1$ of the
Hubble-scale cutoff model of Eq.(\ref{HRCPL}).}
\end{figure}

The  Ricci scalar is given by $R = 6\left(2H^{2} + \dot{H}\right)$.
The corresponding cutoff-scale quantity is $L^{2} = 6/R$, then
\begin{equation}
\rho_H= 3\,c^2\,M_p^2 \,\frac{R}{6} =
\alpha\left(2H^{2} + \dot{H}\right) \ ,  \label{rhR}
\end{equation}
where $\alpha = \frac{3c^{2}}{8\pi G}$. As in the previous
subsection, the balance equation (\ref{cons1}) can be written as
$\dot{\rho}_H + 3H\left(1 + w_{eff} \right)\rho_{H} = 0$ where
\begin{equation}
w_{eff} = \frac{1}{1+r}\left(w + \frac{\dot{w}}{H}\right) = \frac{w
+ \frac{\dot{w}}{H}}{1+r_{0} + 3\left(w - w_{0}\right)} \ .
\label{wef}
\end{equation}
Here $r_{0} = \frac{\Omega_{m0}}{1 - \Omega_{m0}}$. The total
effective EoS parameter is then given by
\begin{equation}
\frac{d\ln H}{d\ln a} = - \frac{3}{2}\frac{1 + r_{0} + 4\left(w -
\frac{3}{4}w_{0}\right)} {1 + r_{0} + 3\left(w - w_{0}\right)} \ ,
\label{dHRpr}
\end{equation}
which can be integrated assuming a form for $w(a)$. Using the CPL
parametrization $w(a)=w_0+(1-a)w_1$ we obtain
\begin{equation}
H(z) = H_0(1 + z)^{\frac{3}{2}\frac{1 + r_0 + \omega_0 +
4\omega_1}{1 + r_0  + 3\omega_1}}\biggr[\frac{1 + r_0 +
3\omega_1\frac{z}{1 + z}}{1 + r_0}\biggl]^{-\frac{1}{2}\frac{1 + r_0
- 3\omega_0}{1 + r_0  + 3\omega_1}}\ . \label{HRCPL}
\end{equation}
The free parameters of this model are $H_{0}$, $r_{0}$, $w_{0}$ and
$w_{1}$. In this case, the minimum value of the four-dimensional
$\chi^{2}$-function is used to determine both $H_{0}$ and $r_{0}$.
Then, a two-dimensional analysis is performed for $w_{0}$ and
$w_{1}$. The results are displayed in Figure \ref{RS1}. The best fit
value we obtained for the Hubble parameter is $h=0.706$ and also
$r_0=0.41$, which translate in a matter density $\Omega_0=0.29$.
Using it, the best fit parameters are $w_{0}=-1.27^{+0.12}_{-0.13}$,
and $w_1=0.99^{+0.30}_{-0.26}$. The $\chi^2_{red} = 1.036$ indicates
this model is among the best fit in this work. In comparison to the
results of \cite{delCampo:2011jp}, our best fit values are
essentially the same at 1$\sigma$.

\subsection{Ricci scale with interaction $Q = 3H\beta\rho_H$.}

\begin{figure}[h!]
\centering \leavevmode\epsfysize=8cm \epsfbox{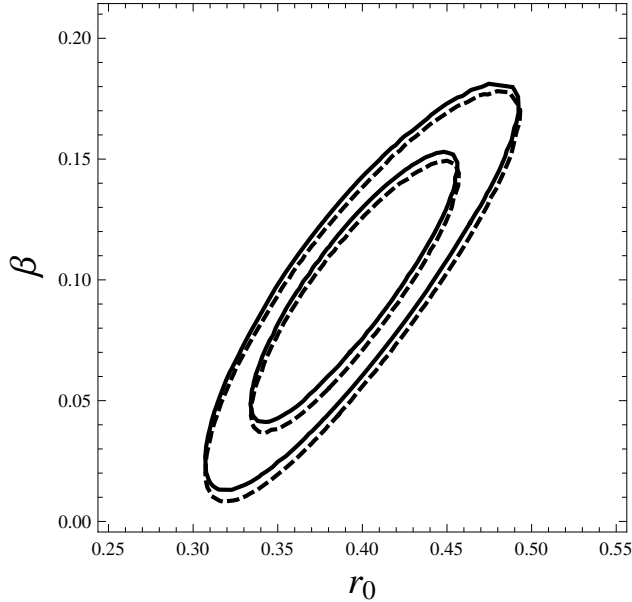}\\
\caption{\label{RS2} Here we display the $68.27\% $ and $95.45\%$
confidence regions for the parameters $r_0$ and $\beta$ in the Ricci
scale with interaction model of Eq.(\ref{HRint}).}
\end{figure}

From the ansatz (\ref{rhR}), in general we can write a relation for
the interaction term
\begin{equation}
Q  = - \frac{3H}{1+r}\left[r w - \frac{\dot{w}}{H}\right]\rho_{H} \
. \label{Q=}
\end{equation}
which is a property of the model. Combining this with
$Q=3H\beta\rho_{H}$, we can get a first-order differential equation
for $w$ which has the solution
\begin{equation}
w = - \frac{1}{6}\frac{u-s - \left(u+s\right)\,A\,a^{s}}{1-A\,a^{s}}
\ , \label{w(a)}
\end{equation}
where
\begin{equation}
u \equiv r_{0} - 3 w_{0} + 3\beta\ ,\quad\ v \equiv r_{0} + 3 w_{0}
+ 3\beta\ ,\quad s\equiv \sqrt{u^{2} - 12\beta\left(1+r_{0} -
3w_{0}\right)} \  \label{uv}
\end{equation}
and
\begin{equation}
A \equiv \frac{v-s}{v+s} \ .  \label{A}
\end{equation}
Using this form (\ref{w(a)}) for $w(a)$ we obtain
\begin{equation}
H(z) = H_0(1 + z)^{\frac{3}{2}\biggr(1 -
\frac{k}{m}\biggl)}\biggr\{\frac{n(1 + z)^{-s} - m}{n -
m}\biggl\}^{\frac{3}{2}\frac{lm - kn}{mns}}\ . \label{HRint}
\end{equation}
Here, one has $H_{0}$, $r_{0}$, $w$ and $\beta$ as free parameters.
We fix $w = - 1$ and determine $H_{0}$ along the lines already
described for the previous cases. The results are displayed in
Fig.(\ref{RS2}). The best fit value we obtained for the Hubble
parameter was $h=0.70$, assuming the value $w=-1$. The best fit
parameters are $r_{0}=0.39^{+0.06}_{-0.06}$, and $\beta =
0.10^{+0.06}_{-0.06}$. The $\chi^2_{red} = 1.033$ shows this
model is one of the best fit in this work, with a negligible statistical
difference with respect to the $\Lambda$CDM (See Table I). In
comparison to the results of \cite{delCampo:2011jp}, our best fit
values are essentially the same at 1$\sigma$.

As a summary of our results we display the $\chi^2_{min}$, and the
Akaike Information Criteria (AIC) and Bayesian Information Criteria
(BIC) for each model, in comparison with the $\Lambda$CDM model in
Table I. Although the $\Lambda$CDM model is the best fit to the
data, considering the $\chi^2_{min}$ values, the Future Event
Horizon model with $\xi=2$ and both Ricci scale models (with CPL and
interaction), appears in reasonable statistical agreement with the
data. However, considering the number of free parameters of each
model, using for example the BIC criteria \cite{bicpaper}, neither
of these models are really competitive with the $\Lambda$CDM model.

\begin{table}[!h]
\begin{center}
\begin{tabular}{|p{1.7cm}|p{1.7cm}|p{1.7cm}|p{1.7cm}|p{1.7cm}|p{1.7cm}|p{1.7cm}|p{1.7cm}|}
\hline
 & $\Lambda CDM$ & Hubble Radius & Future ($\xi=1$) & Future ($\xi=2$) & Future ($\xi=3$) & Ricci CPL & Ricci with Q \\
\hline
$\chi^2_{min}$& 675.57 & 794.37 & 710.01 & 677.14 & 689.34 & 677.51 & 675.28 \\
\hline
$\chi^2_{red}$& 1.030 & 1.213 & 1.084 & 1.034 & 1.052 & 1.036 & 1.033 \\
\hline
$\triangle$AIC& 0 & 121 & 36 & 3.6 & 16 & 3.9 & 3.7 \\
\hline
$\triangle$BIC& 0 & 125 & 41 & 8.1 & 20 & 15 & 13 \\
\hline
\end{tabular}
\caption{This table is a summary of the statistical analysis using
all the data; 557 SNIa, 70 Strong Lensing points, 28 Hubble function
points, and three BAO scale points. We show both the $\chi^2_{min}$
and the reduced $\chi^2_{red}$, which takes into account the number
of free parameters to fit the data. We observe that, although the
$\chi^2_{min}$ values for the Future Event Horizon with $\xi=2 $,
and the Ricci scale with interaction model, show a fit similar to
the $\Lambda$CDM,  taking into account the number of free parameters
for each model, through the AIC and BIC criteria, the data suggest
neither of the holographic models is competitive with the
$\Lambda$CDM model.}
\end{center}
\label{tab1}
\end{table}

\section{Conclusion}

In this work we have used strongly gravitationally lensed (SGL)
systems, to provide additional constraints on a set of holographic
dark energy models previously considered in \cite{delCampo:2011jp}.
In addition to the SGL data, in this paper we have used the largest
set of measurements of the Hubble parameter $H(z)$ in the range of
redshifts $0.07 \leq z \leq 2.3$ \cite{Farooq:2013hq}, recent data
from BAO \cite{Blake:2011en} and supernovae \cite{Union2}.

We found that the $\Lambda$CDM, with two free parameters $\Omega_m$
and $h$, is the best fit to all  the data. Although the statistical
comparison among $\chi^{2}_{min}$ values seems to indicate that the
Future Event Horizon with $\xi=2 $, and the Ricci scale  holographic
model, show interesting agreement  with the observations, using a
stringent test, using the AIC and BIC criteria, lead us to the
result that neither of the holographic models are competitive with
the $\Lambda$CDM.

These results show the importance of Strong Lensing measurements to
provide additional observational constraints to alternative
cosmological models, which are necessary to shed some light into the
dark universe.

\begin{acknowledgments}
This work was funded by Comision Nacional de Ciencias y
Tecnolog\'{\i}a through FONDECYT Grants 1110230 (SdC and VHC),
1120741 (AB,VM), and also from DI-PUCV Grants 123.710 (SdC) and
through DIUV grant 13/2009 (VHC).
\end{acknowledgments}

\end{document}